# Weight Measurements of High-Temperature Superconductors during Phase Transition in Stationary, Non-Stationary Condition and under ELF Radiation

Martin Tajmar[1], Klaus Hense[1], Klaus Marhold[1], Clovis de Matos[2]

[1] *Space Propulsion, ARC Seibersdorf research, A-2444 Seibersdorf, Austria*
[2] *ESA-HQ, F-75015 Paris, France*
*+43-50550-3142; martin.tajmar@arcs.ac.at*

**Abstract.** There have been a number of claims in the literature about gravity shielding effects of superconductors and more recently on the weight reduction of superconductors passing through their critical temperature. We report several experiments to test the weight of superconductors under various conditions. First, we report tests on the weight of YBCO and BSCCO high temperature superconductors passing through their critical temperature. No anomaly was found within the equipment accuracy ruling out claimed anomalies by Rounds and Reiss. Our experiments extend the accuracy of previous measurements by two orders of magnitude. In addition, for the first time, the weight of a rotating YBCO superconductor was measured. Also in this case, no weight anomaly could be seen within the accuracy of the equipment used. In addition, also weight measurements of a BSCCO superconductor subjected to extremely-low-frequency (ELF) radiation have been done to test a claim of weight reduction under these conditions by De Aquino, and again, no unusual behavior was found. These measurements put new important boundaries on any inertial effect connected with superconductivity - and consequently on possible space related applications.

## INTRODUCTION

Since the publications of Podkletnov (Podkletnov and Nieminen, 1992, Podkletnov, 1997, 2003), there has been much debate in the literature (De Podesta and Bull, 1995, Unnikrishnan, 1996), about his claim of reducing a sample weight by 0.03 – 2.1% placed over a levitating YBCO high temperature superconductor (non-rotating and rotating using field coils). Several replication experiments were carried out to date, and, although not all conditions reported by Podkletnov were fully reached in some cases (e.g. size of superconductor or frequency of field coil), no successful replication of his claims were reported (Li et al., 1997; Woods et al., 2001; Hathaway, Cleveland and Bao, 2003).

More recently, claims about the change of mass from YBCO and BSCCO high temperature superconductors were reported in the range of 0.05% and 0.5% respectively passing through their critical temperature (Rounds, 1997; Reiss, 1999).

Rounds used a styropor dewar placed above a microbalance. Inside the dewar, which was filled with liquid nitrogen, an YBCO sample was placed above a permanent magnet. He then monitored the weight of the dewar assembly until the nitrogen was fully evaporated; hence he observed the superconductor passing from the superconductive to the normal phase. After analyzing the weight slope and comparing it with a reference sample slope, he claimed to have observed steps around the critical temperature of 90 K showing a sudden weight change of the YBCO of about 0.05% from its weight.

Reiss used a PTFE capsule filled with BSCCO superconductors of 2212 type alone or mixing them with permanent magnets. This capsule, connected to a microbalance, was then immersed into a dewar filled with liquid nitrogen, which was slowly evaporating. After analyzing the weight slope of the superconductors with respect to measurements done using a non-superconductive reference sample (PVC), Reiss claimed that the superconductors show an increase in weight of about 0.5% in their superconductive state with respect to their non-superconductive state.

Woods et al performed an experiment similar to the one by Rounds ruling out any anomalous behavior down to 0.04% of the superconductor’s weight. His conclusion was that Rounds has been looking for gravitational anomalies at temperatures well below the critical temperature.

Recently, De Aquino developed a theory and published experimental results claiming that an Hg superconductor lost weight when subjected to Extremely-Low-Frequency (ELF) radiation. His experimental results matched exactly his theoretical predictions (De Aquino, 2002 and 2004).

In this paper presented are the weight measurements of YBCO and BSCCO superconductors during their phase transition with an improvement by more than two orders of magnitude with respect to the previous measurements. In addition, for the first time, we also report weight measurements for YBCO under non-stationary conditions (rotating and accelerating). No weight anomalies were seen within the resolution of the experiments. We also tested the weight reduction claim by De Aquino and put an BSCCO superconductor under ELF radiation. No weight anomalies were seen in this case as well up to the resolution of the equipment used. These results put new bounds on claimed anomalies of superconductor masses as well as on ongoing theoretical studies on extending the equivalence principles to quantum materials (Chiao, 2002). All samples used for the measurements are listed in Table 1. The YBCO, aluminum and stainless steel samples were disc shaped whereas the BSCCO sample was ring shaped.

**TABLE 1.** Summary of Samples used in Measurements.

| Type | Weight | Dimensions | Critical Temperature |
|---|---|---|---|
| $YB_2C_3O_{7-x}$ (YBCO) | 70.3653 g | ∅ 36 mm, height 11.25 mm | 90 K |
| Aluminum | 30.9525 g | ∅ 36 mm, height 11.25 mm | - |
| Stainless Steel (1.4301) | 90.3306 g | ∅ 36 mm, height 11.25 mm | - |
| $Bi_{1.8}Pb_{0.26}Sr_2Ca_2Cu_3O_{10-x}$ (BSCCO) | 30.1277 g | Outer Diameter: 47.65 mm<br>Wall Thickness: 3.76 mm<br>Height: 14.7 mm | 108 K |
| $Bi_{1.8}Pb_{0.26}Sr_2Ca_2Cu_3O_{10-x}$ (BSCCO) | 188.83 g | Outer Diameter: 47.65 mm<br>Wall Thickness: 3.76 mm<br>Height: 60.69 mm | 108 K |

## WEIGHT MEASUREMENTS FOR STATIONARY HIGH-TEMPERATURE SUPERCONDUCTORS – SETUP A

For the first measurements, we intended to closely resemble the simple setup from Rounds (see Fig. 1). For the weight measurement we used a Sartorius BP 110 S balance with a maximum weight capability of 110 g and a resolution of 0.1 mg. The balance was connected via a serial interface to a PC. Data acquisition was done using a custom LabView program. On top of the balance was a plastic cup thermally isolated from the balance by a block of styropor. The samples were put in the middle of the cup, which was then filled with liquid nitrogen and closed on top by a sheet of paper. Due to the lack of thermal isolation from the environment, nitrogen evaporation was very fast. In all tests, the cup was usually empty within about 10 minutes. To get an approximation of the time required to pass through the superconductor’s critical temperature, a permanent magnet was placed above the YBCO which was levitating when the superconductor was immersed in the liquid nitrogen. After all nitrogen evaporated, the magnet continued to levitate for about 30 s. Therefore we expected to see possible weight anomalies at about this time scale after the complete evaporation of nitrogen from the plastic cup.

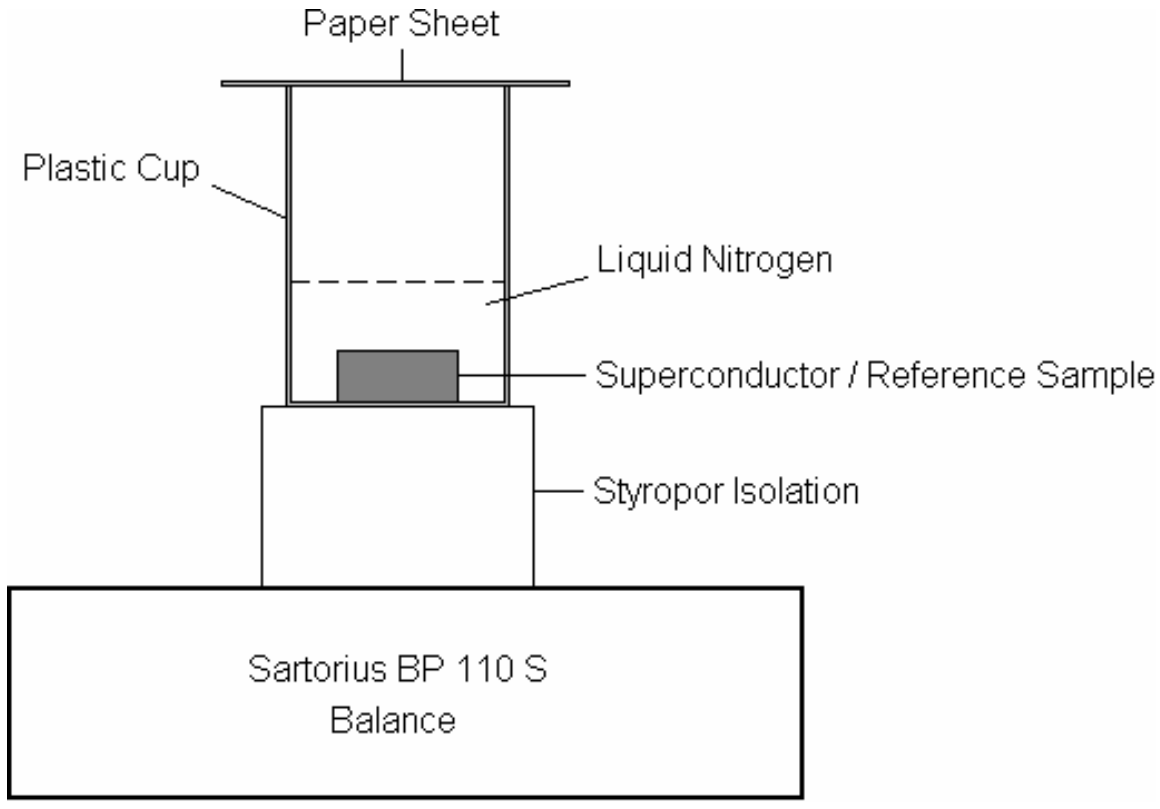

**FIGURE 1.** Weight Measurement Setup A.

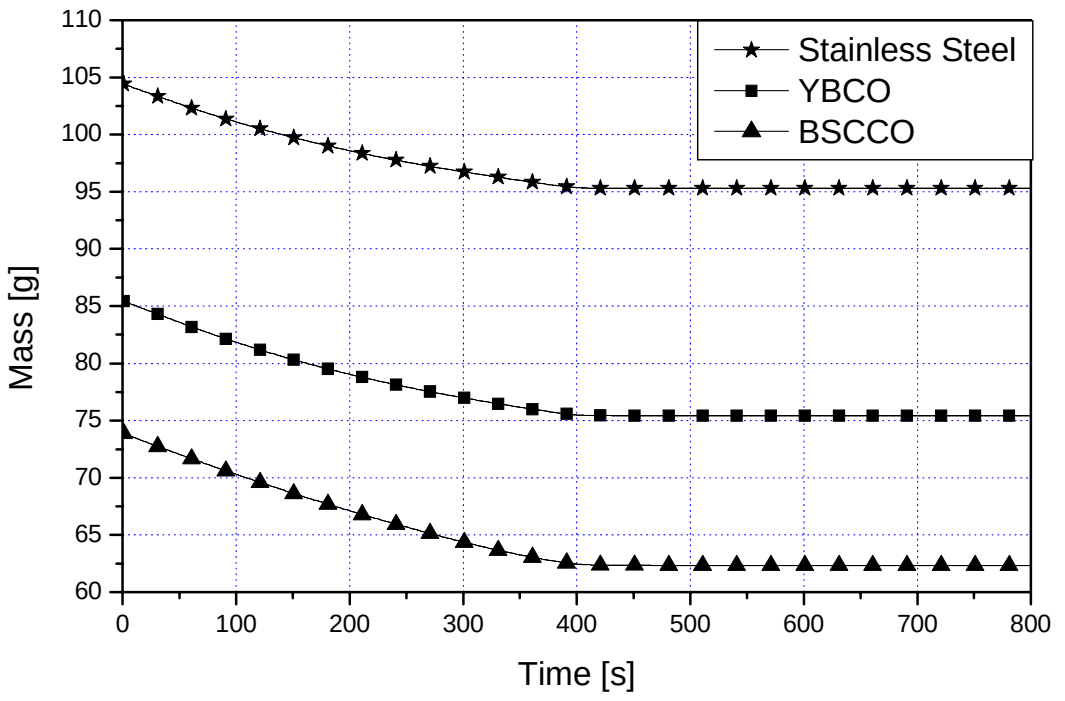

**FIGURE 2.** Superconductor Weight Measurements in Setup A.

Fig. 2 shows the weight-time dependence for YBCO and BSCOO superconductors as well as for a stainless steel reference sample. Nitrogen evaporated until 400 s, after that we expected to see possible weight anomalies. Analyzing the weight curves at high resolution, no steps other than the resolution of the balance readings can be seen from the 400 s down to 800 s – in a time frame where the superconductor must have passed its critical temperature.

In order to resemble Round's measurement best, a Rare-Earth permanent magnet with a surface field strength of about 1 T was mechanically attached to the YBCO and put in the middle of the plastic cup. This traps a considerable magnetic field once the YBCO becomes superconductive. Also in this case, no weight anomaly has been seen.

Summarizing the measurements in Setup A, no weight anomaly was seen for an YBCO superconductor (with and without trapped magnetic field) down to 0.00014% and for a BSCCO superconductor down to 0.00033% of their respective weight passing through their critical temperatures according to the resolution of our balance.

## WEIGHT MEASUREMENTS FOR STATIONARY HIGH-TEMPERATURE SUPERCONDUCTORS – SETUP B

This setup was intended to allow for a much slower heating and to monitor the temperature of the sample tested. A sketch is shown in Fig. 3 which is very similar to the setup used by Reiss for his measurements. In Setup B, the same Sartorius BP 110 S balance with a resolution of 0.1 mg was used. The balance was fixed above a glass cryostat from which the samples were hanging down into the cryostat using a thin Ni wire with a diameter of 125 µm. Next to the samples, a PT100 platinum resistor was mounted and connected to the PC using the 4-wire method and a constant current of 1 mA. The resistor did not touch the sample (in order to avoid influence due to e.g. ice formation on the wires creating noise on the weight measurement) but was in close vicinity (about 1 cm) to have reasonable temperature accuracy. A particular problem was noted from the resistor when it came out of the liquid. Due to the electrical power that was fed into the sensor (200 mW), the temperature of the gas surrounding the resistor was increasing by about 40 K. This also changed the buoyancy of the samples floating in the nitrogen gas and caused an apparent weight change. Therefore, the PT100 was mounted slightly above the superconductor so that it would come out first and not spoil the superconductor weight measurements. The temperature slope was then corrected by knowing that when the superconductor is touching the liquid only with its base, its temperature would be still the temperature of the liquid nitrogen, i.e. 77 K. After the cryostat was filled with liquid nitrogen (according to optical observation, the liquid was always about 10 cm above the sample), a foam was put on top of the cryostat to reduce in-flow of humid air.

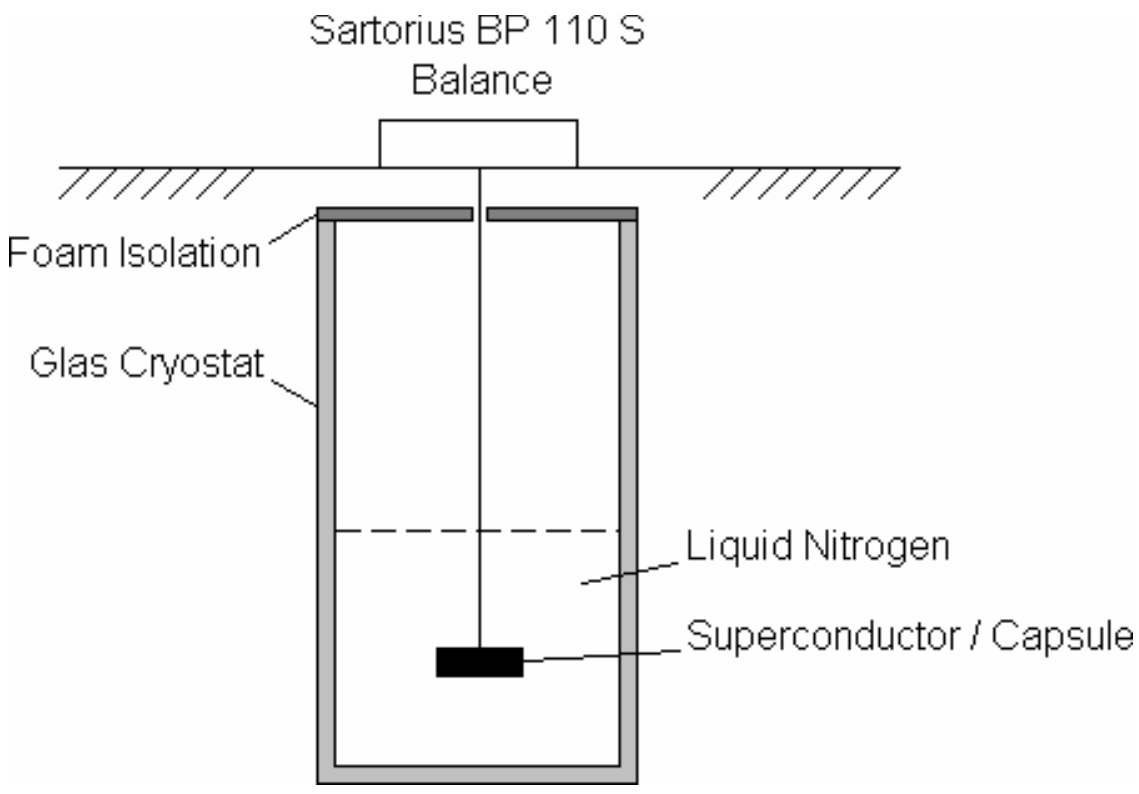

**FIGURE 3.** Weight Measurement Setup B.

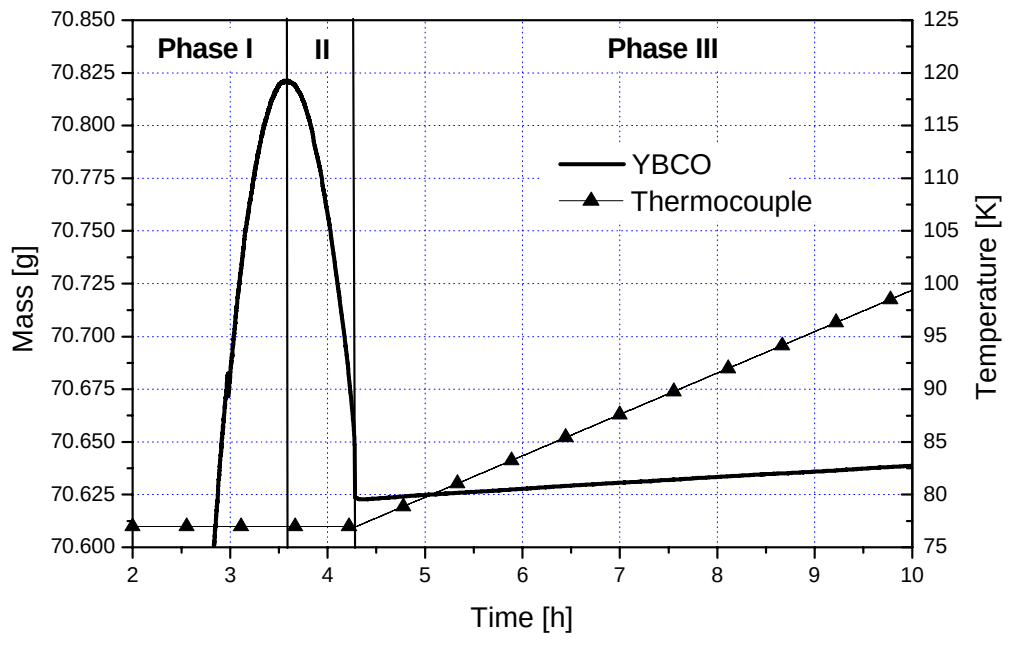

**FIGURE 4.** YBCO Weight Measurement and Temperature Reading in Setup B (Phase I … Buoyancy in Liquid Nitrogen, Phase II … Liquid Nitrogen Droplets Evaporate on Sample, Phase III … Buoyancy in Nitrogen Gas).

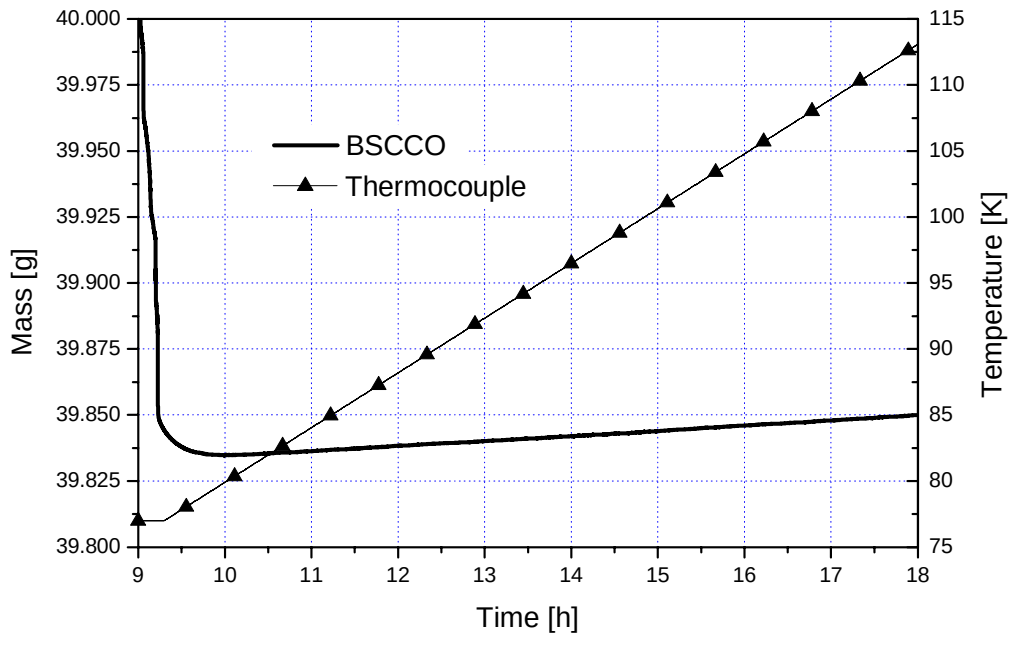

**FIGURE 5.** BSCCO Weight Measurement and Temperature Reading in Setup B.

Fig. 4 shows the weight and temperature curve for an YBCO sample. Three phases are visible in the plot: In Phase I, the YBCO is still immersed in the liquid nitrogen and is slowly coming out. Due to the decrease of buoyancy (the volume of the sample in the liquid nitrogen gets smaller as it is coming out), the apparent weight is going up until the sample is only touching the liquid at its base. In Phase II, liquid nitrogen droplets on the base of the sample are evaporating corresponding to an apparent weight decrease. In Phase III, the sample is free of liquid nitrogen and floating only in the nitrogen gas phase. The overpressure of nitrogen in the cryostat limits air (oxygen and humidity) from flowing in and therefore reduces water ice formation significantly (it was observed that when all nitrogen was vented out of the cryostat, the weight of the sample increased drastically due to water ice formation). As the nitrogen gas gets warmer, the buoyancy of the sample is changing too. Therefore, the weight slope is slightly increasing.

According to the temperature reading, the critical temperature of the YBCO (about 90 K) should have been reached 7-8 hours after the start of the measurement. However, no anomalous change in the weight slope could be detected within the resolution of the balance. A similar test was done using the BSCCO sample, the gas phase is shown in Fig. 5. Due to the higher critical temperature of the BSCCO (about 108 K), we expected weight anomalies between 16 and 17 hours after the start of the experiment. Also in this case, no change in the weight slope could be seen.

In order to get a better temperature measurement at the superconductor sample, a capsule was built out of aluminum to which a PT100 platinum resistor for the temperature measurement was directly mounted and which hosted the superconductor or reference sample (see Fig. 6). Fig. 10 shows a plot of the weight measurement for the YBCO and the aluminum reference sample (with an offset of 37.9341 g to account for the lower density of aluminum for better comparison) versus the measured temperature. The peaks from 80 – 85 K are due to the more complicated outer surface of the capsule and the resulting different evaporation of droplets and buoyancy. The slightly different weight slope from the reference sample with respect to the YBCO is due to the difference in specific heat. Again, no anomaly can be seen around the critical temperature from YBCO at 90 K. The procedure was also repeated attaching a Rare Earth permanent magnet on the bottom of the capsule (see Fig. 7). Also in this case, no anomaly was seen within the resolution of the balance.

Summarizing the measurements in Setup B, the result is similar to the one in Setup A, as there was no weight anomaly seen for an YBCO superconductor (with and without trapped magnetic field) down to 0.00014% and for a BSCCO superconductor down to 0.00033% of their respective weight according to the resolution of our balance.

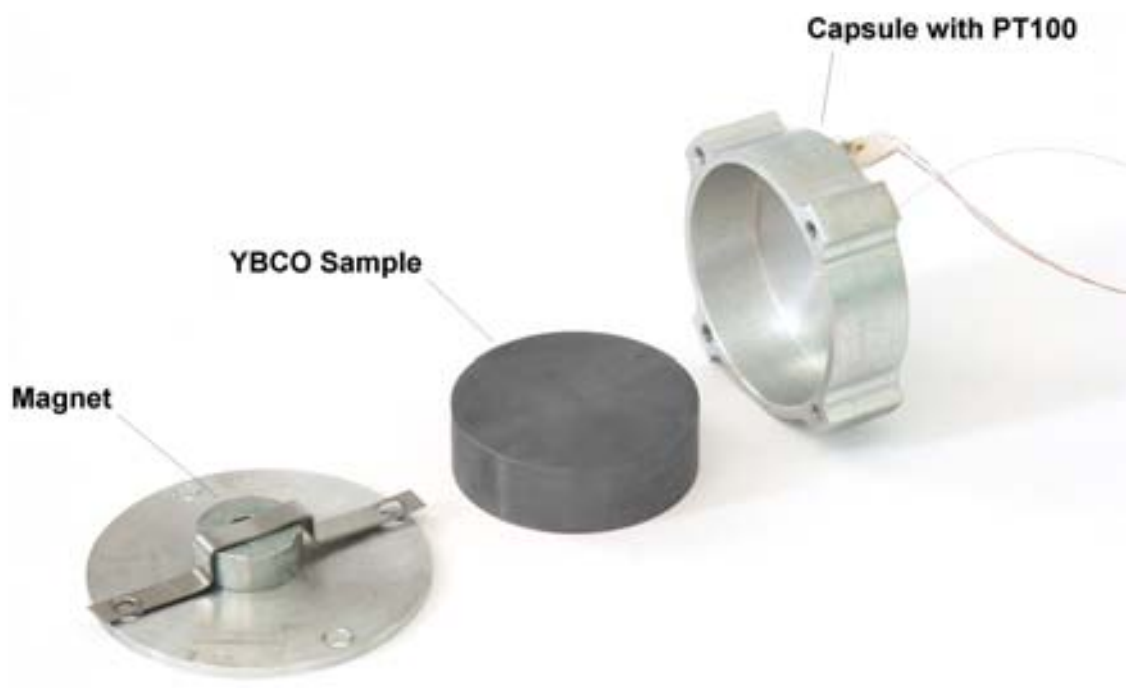

**FIGURE 6.** Capsule for Setup B (From Left to Right: Bottom Plate with Optional Magnet, YBCO Sample and Capsule Main Part with PT100 Resistor Attached).

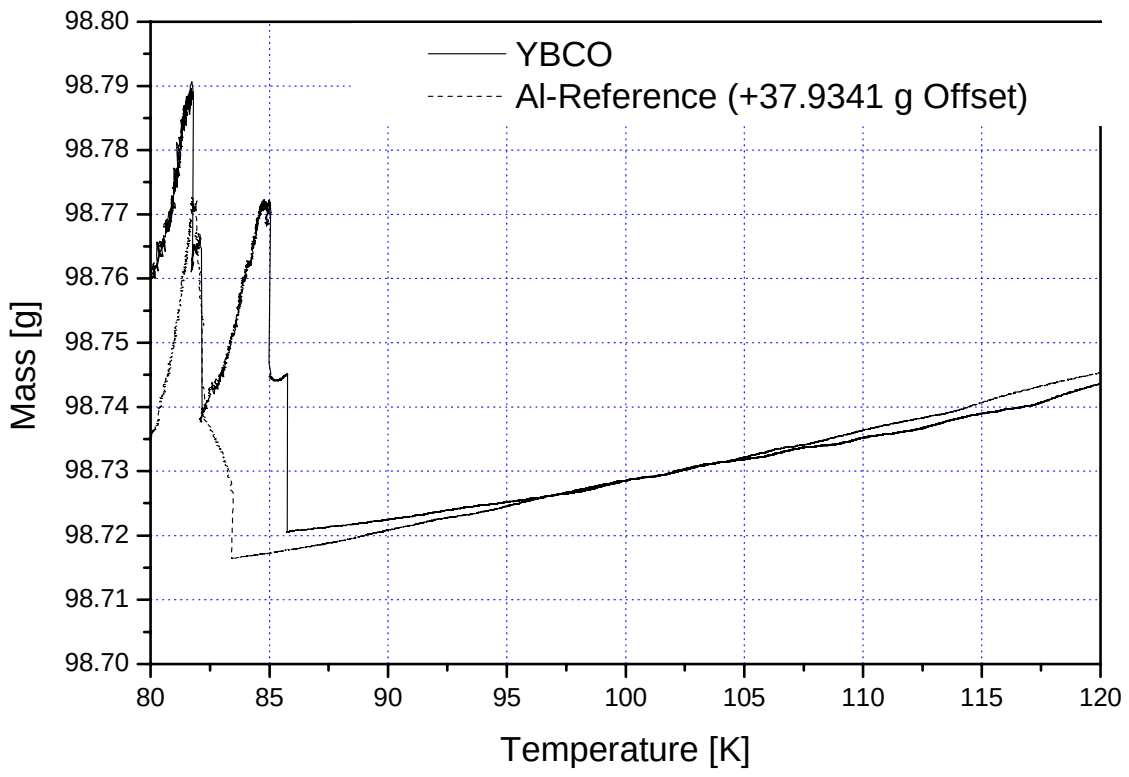

**FIGURE 7.** YBCO and Aluminum Reference in Capsule Weight Measurement against Temperature Reading in Setup B (Offset of 37.9341 g for Aluminum to Compensate for Lower Density with Respect to YBCO for Better Comparison).

## WEIGHT MEASUREMENTS FOR NON-STATIONARY HIGH-TEMPERATURE SUPERCONDUCTOR – SETUP C

Next, we tried to see if rotation of the sample could cause an anomalous weight effect. Therefore we built an insert with an electric motor (4.5 – 15 V DC with a reduction gear of 11:1 to achieve a maximum angular velocity of 1800 rpm at maximum voltage) as shown in Fig. 8. The actual speed was determined with an optical encoder. The superconductor and reference sample were again put into a capsule. This capsule was thermally isolated from room temperature with several radiation shields and styropor sheets. A picture of the complete insert is shown in Fig. 9. As this setup is much heavier, a different balance had to be used for the measurements. We selected the Sartorius MC 1 with a total weight capacity of 2.2 kg and a resolution of 10 mg. Both balance and speed measurement as well as drive commands for the motor were implemented in a LabView program.

As we already tested the weight of a superconductor for stationary conditions, during the non-stationary tests, the samples were always immersed within the liquid (we knew that for the start of the experiment at angular velocity = 0 no weight anomaly was to be expected). Within 2 seconds we then accelerated the sample to maximum speed at 1800 rpm, remaining at this speed for about 30 seconds, and then stopped rotation. Fig. 10 shows the speed measurement and the balance reading for an YBCO sample. The main problem during this experiment was the stirring of the liquid nitrogen by the sample. Due to the viscosity, a parabolic surface was formed, bringing the liquid nitrogen in contact with warmer parts of the reservoir, which results in a rise of the evaporation rate that is causing the change in the weight slope. This increased evaporation rate is by far stronger than any observed effect. The slope in Fig. 10 is corresponding to a rotational speed of 1800 rpm. A different rotational speed would also change this slope. A weight anomaly was to be expected by a shift of the curve at the beginning and at the end of the rotation. In fact, there are two peaks at these positions, however both showing in the same direction. Moreover, a similar test with an aluminum reference sample showed the same peaks and weight slope with similar values.

Therefore, no weight anomaly could be detected for an YBCO up to a resolution of about 100 mg. This translates to a weight percentage of 0.14% of the sample's weight.

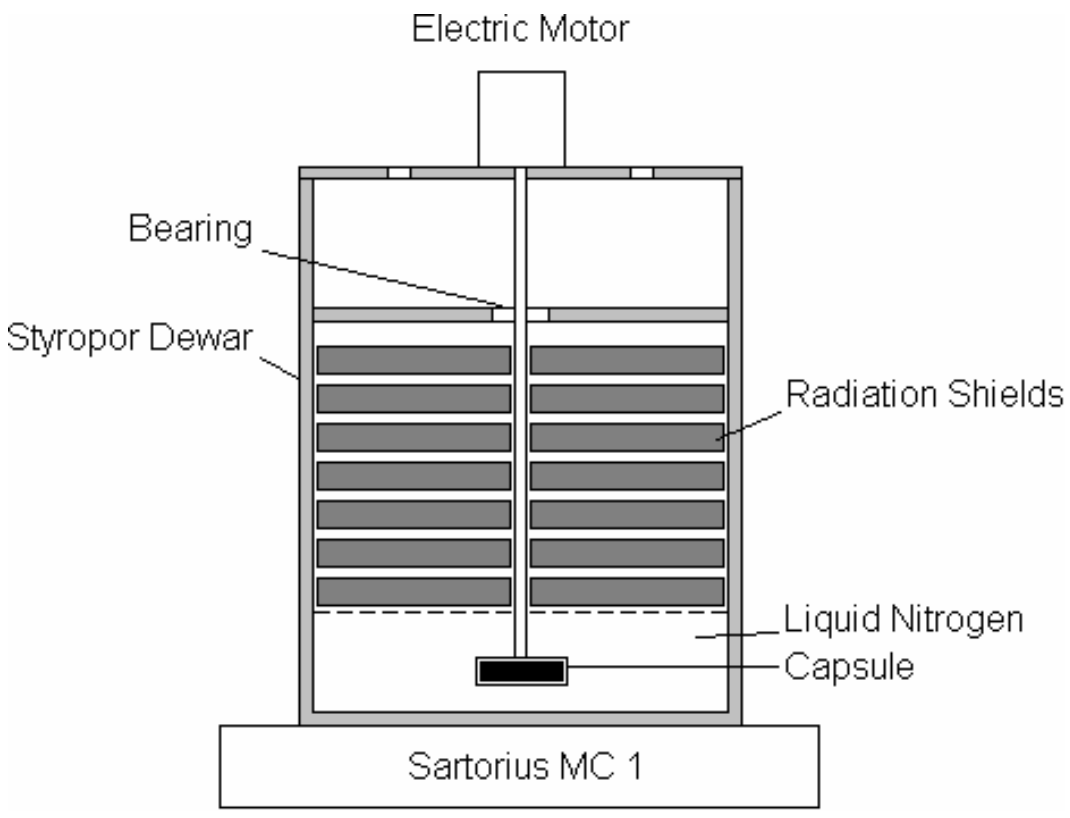

**FIGURE 8.** Weight Measurement Setup C.

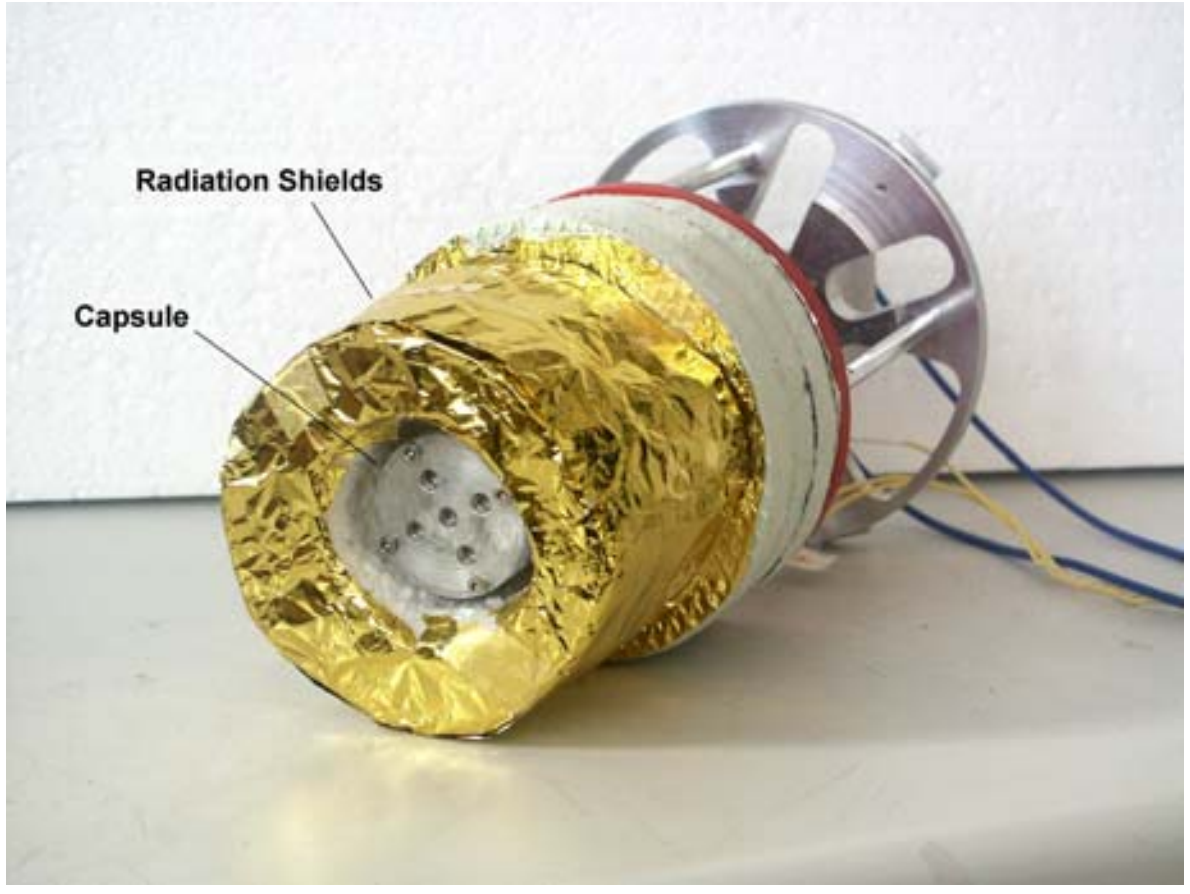

**FIGURE 9.** Insert for Dewar in Setup C (Capsule with YBCO Surrounded by Radiation Shields).

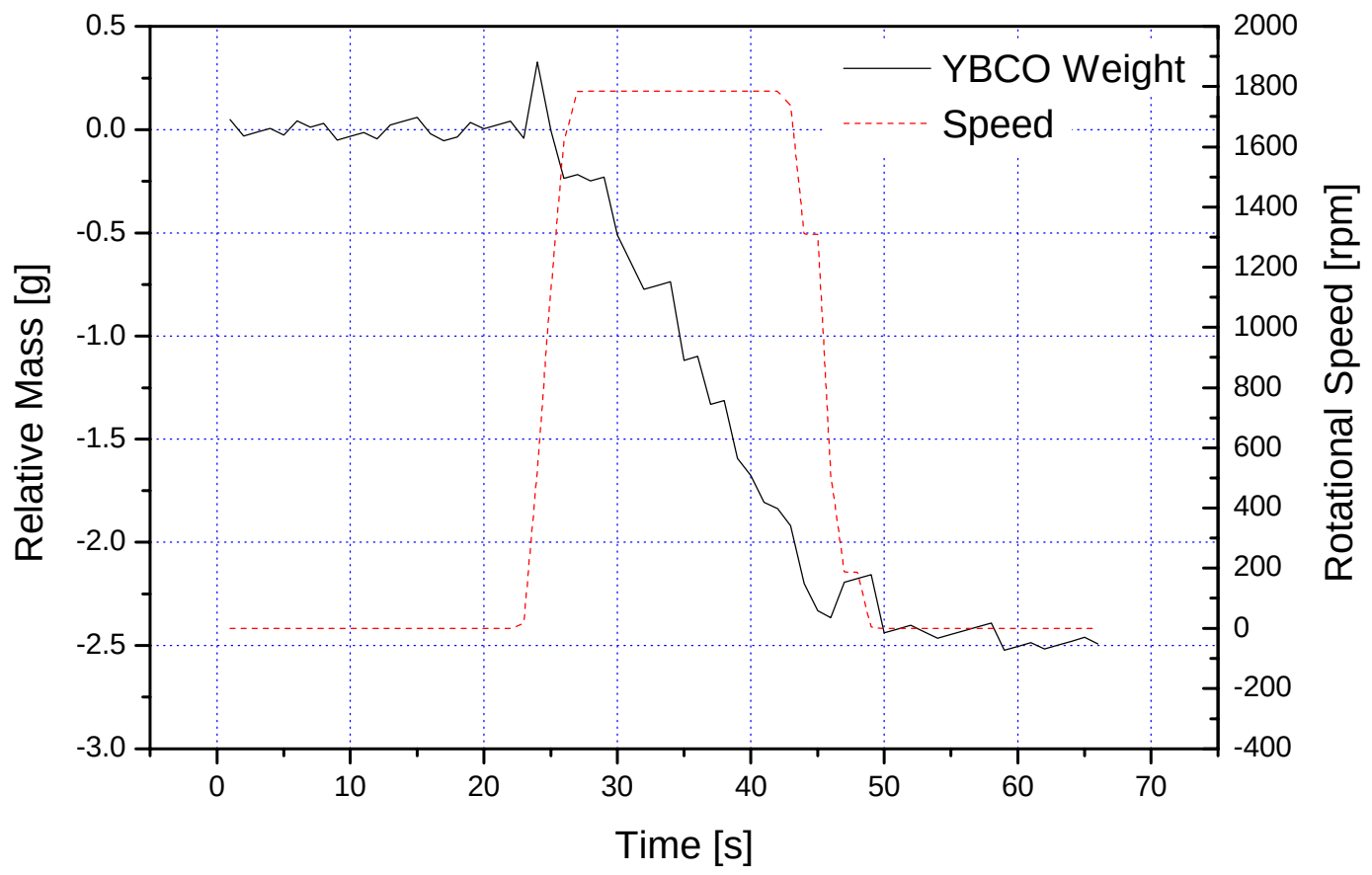

**FIGURE 10.** Relative Weight Measurement of YBCO under Rotation in Setup C.

## WEIGHT MEASUREMENTS FOR STATIONARY HIGH-TEMPERATURE SUPERCONDUCTORS UNDER ELF RADIATION – SETUP D

Finally, we assessed the claim from De Aquino if a superconductor looses weight when subjected to Extremely-Low-Frequency (ELF) radiation. According to his analysis, the mass of a superconductor $m_{SC}$ should change according to (De Aquino, 2002, 2004):

$$m'_{SC} = m_{SC} - 2n_s V_{SC} m_e \cdot \left[ \sqrt{1 + \left( \frac{E}{m_e c} \cdot \frac{q}{f} \right)^2} - 1 \right] . \tag{1}$$

For this test, we put a BSCCO ring (see last line in Table 2) inside a cylindrical capacitor with $r_o$=56 mm and $r_i$=32 mm. In this case, the electric field is given by

$$E = U \cdot \left[ \frac{r_o + r_i}{2} \cdot \ln\left( \frac{r_o}{r_i} \right) \right]^{-1} . \tag{2}$$

For BSCCO, the Cooper pair density at the temperature of the liquid nitrogen is about $n_s$=5.3x10$^{26}$ m$^{-3}$, and the volume of the ring was $V_{SC}$=3.45x10$^{-5}$ m$^3$. For our initial weight of 188.83 g, we would have then expected to see a weight variation ranging from 0.04 % up to 4.2 % when subjected to an ELF radiation with amplitude ranging from 1-10 V and frequency from 1-10 Hz.

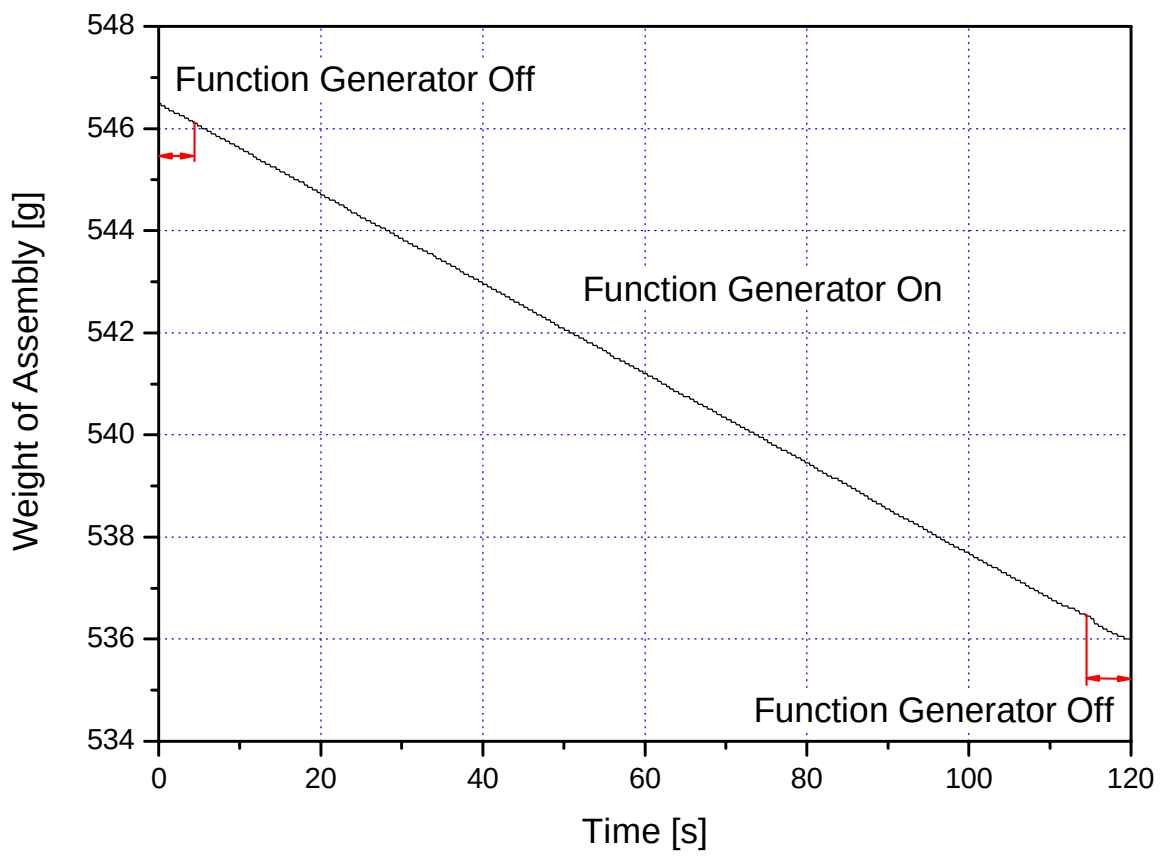

**FIGURE 11.** Weight Measurement of BSCCO Assembly under ELF Radiation with an Amplitude Varying from 1-10 V at 3 Hz – Setup D.

The superconductor and the capacitor were put on the bottom of a styropor dewar that was filled with liquid nitrogen. The whole assembly was then put on a Sartorius MC 1 balance similar to the one used in Setup C. Fig. 11 shows the weight of the complete assembly versus time for the case of 3 Hz (the constant weight loss is caused by the evaporation of liquid nitrogen). At the left and right side of the plot, the function generator was shut off for calibration purposes as indicated. Between the calibration periods, the function generator was on and a rectangular shaped signal with amplitude from 1 up to 10 V at a frequency of 3 Hz was applied. No difference in the weight slope between function generator on and off was seen. This was also verified for other frequency ranges from 1 up to 10 Hz.

Therefore, no weight anomaly could be detected for a BSCCO up to a resolution of 50 mg. This translates to a weight percentage of 0.026% of the sample's weight.

**TABLE 2.** Summary of Measurements Results.

| **Type** | **Stationary Condition** | **Non-Stationary Condition** | **Weight Anomaly Ruled Out Upper Limit**[*] | **Previous Claim** |
|---|---|---|---|---|
| YBCO | Passing through $T_c$ With/Without Magnet | | 0.00014 % | > 0.05 % |
| YBCO | | Rotating at 1800 rpm | 0.14 % | |
| | | Acceleration to 1800 rpm in 2 seconds | 0.14 % | |
| BSCCO | Passing through $T_c$ | | 0.00033 % | 0.5 % |
| BSCCO | Subjected to ELF Radiation Amplitude: 0-10 V Frequency: 0-10 Hz | | 0.026 % | 0.04 – 4.2 % |

[*]Percentage with Respect to Weight of Sample

## CONCLUSIONS

The weight of YBCO and BSCCO high temperature superconductors has been measured while passing through their respective critical temperatures. Also the weight of an YBCO under rotation up to 1800 rpm was evaluated as well as the weight of a BSCCO under ELF radiation. No weight anomaly could be identified up to the resolution of the used equipment. The results are summarized in Table 2. Within the reached measurement accuracy, our results rule out previous claims by Rounds, Reiss and De Aquino on the mass of superconductors.

Up to the knowledge of the authors, no other measurements than the ones presented and the ones included in the references attached to this paper have been conducted on the weight of a material between its superconductive and normal-conductive phase. The present paper shall establish first experimental bounds on this value – which is one of the most fundamental material's properties. Our results were obtained using a rather simple setup – hence parasite effects such as buoyancy or moisture condensation were present which limited the achievable accuracy. A more in-depth study would require a vacuum balance and a temperature-controllable cryostat to get rid of the parasite effects encountered in these first measurements as well as to investigate the mass ratio between a material's superconductive and normal conductive phase to an even higher accuracy. Similar experiments with superfluids instead of superconductors should be carried out to cover completely the issue of the weight of quantum materials at their respective transition phase.

## NOMENCLATURE

$m_{SC}$ = mass of superconductor (kg)
$m_e$ = mass of electron (kg)
$n_s$ = density of free electroncs or Cooper-pairs ($m^{-3}$)
c = speed of light ($3x10^8$ $m.s^{-1}$)
E = electric field ($V.m^{-1}$)
f = frequency (Hz)
q = electric charge ($1.6x10^{-19}$ C)
$r_o$ = outer radius (m)
$r_i$ = inner radius (m)
U = applied voltage (V)